\documentclass[aps,prl,twocolumn]{revtex4-1}
\usepackage{graphicx}
\usepackage{dcolumn}
\usepackage{bm}
\usepackage{multirow}
\usepackage{ulem}
\usepackage{color}
\usepackage{subfigure}
\usepackage{amsmath}
\usepackage{upgreek}

\begin{document}
\title{Highly charged Nd$^{9+}$ Ion: A potential candidate of $\upmu$Hz linewidth optical clocks for probing fundamental physics}

\author{$^{1,6}$Yan-mei Yu}
\email{ymyu@aphy.iphy.ac.cn}
\author{$^2$Duo Pan}
\email{panduo@pku.edu.cn}
\author{$^{3,4}$Shaolong Chen}
\author{$^5$Bindiya Arora}

\author{$^{3,4}$Hua Guan}
\email{guanhua@apm.ac.cn}
\author{$^{3,4}$Kelin Gao}
\author{$^2$Jingbiao Chen}

\affiliation{$^1$Beijing National Laboratory for Condensed Matter Physics, Institute of Physics, Chinese Academy of Sciences, Beijing 100190, China}
\affiliation{$^2$State Key Laboratory of Advanced Optical Communication Systems and Networks, Department of Electronics, Peking University, Beijing 100871, China}
\affiliation{$^3$State Key Laboratory of Magnetic Resonance and Atomic and Molecular Physics, Innovation Academy for Precision Measurement Science and Technology, Chinese Academy of Sciences, Wuhan 430071, China}
\affiliation{$^4$Key Laboratory of Atomic Frequency Standards, Innovation Academy for Precision Measurement Science and Technology, Chinese Academy of Sciences, Wuhan 430071, China}
\affiliation{$^5$Department of Physics, Guru Nanak Dev University, Amritsar, Punjab 143005, India}
\affiliation{$^{6}$University of Chinese Academy of Sciences, 100049 Beijing, China}

\begin{abstract}
An active optical clock based on highly charged Nd$^{9+}$ ion is proposed for the first time.  The clock can offer ultra-narrow linewidth at the $\upmu$Hz-level which is more than two-order of magnitude below the currently recorded laser linewidth. Operating at 605(90) nm superradiation lasing transition between the $5p^2~4f$ ground state and one of the long-lived $5p~4f^2$ excited state, the proposed active clock  is inherently immune against the cavity noise which provides high stability. The clock serves as a sensitive probe with high sensitivity to variation of the fine-structure constant with accuracies below 10$^{-19}$ level. Sophisticated relativistic many-body methods are employed to predict related atomic properties that have corroborated the above findings.
\end{abstract}
\date{\today}

\maketitle

\begin{figure}[t]
\centering
 \subfigure[ ]{
\includegraphics[width=6cm,height=5.5cm]{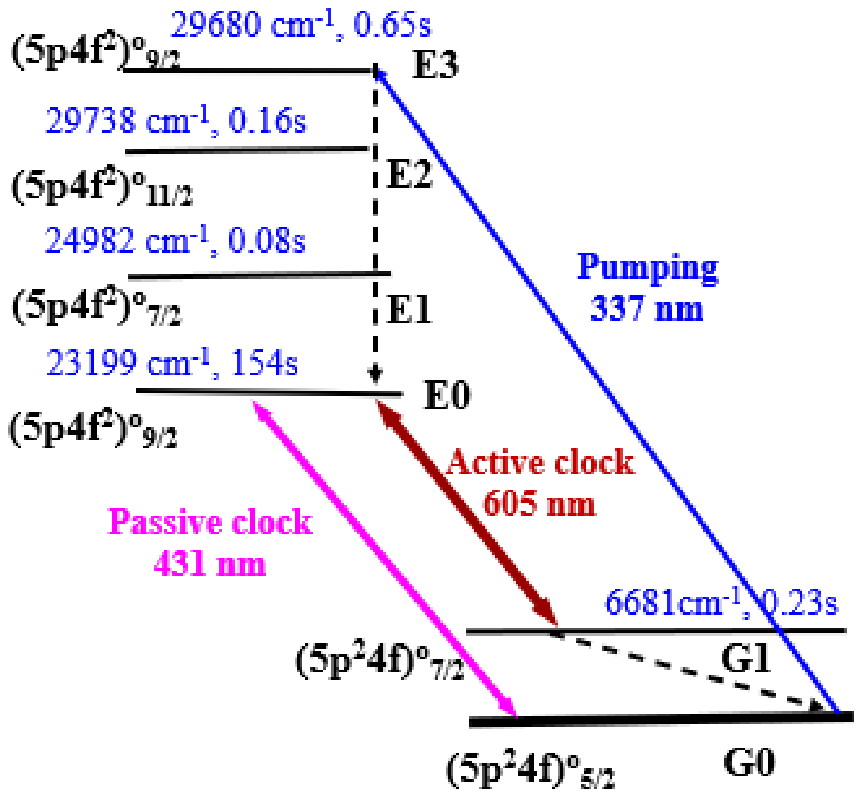}
}
 \subfigure[ ]{
\includegraphics[width=6cm,height=6.0cm]{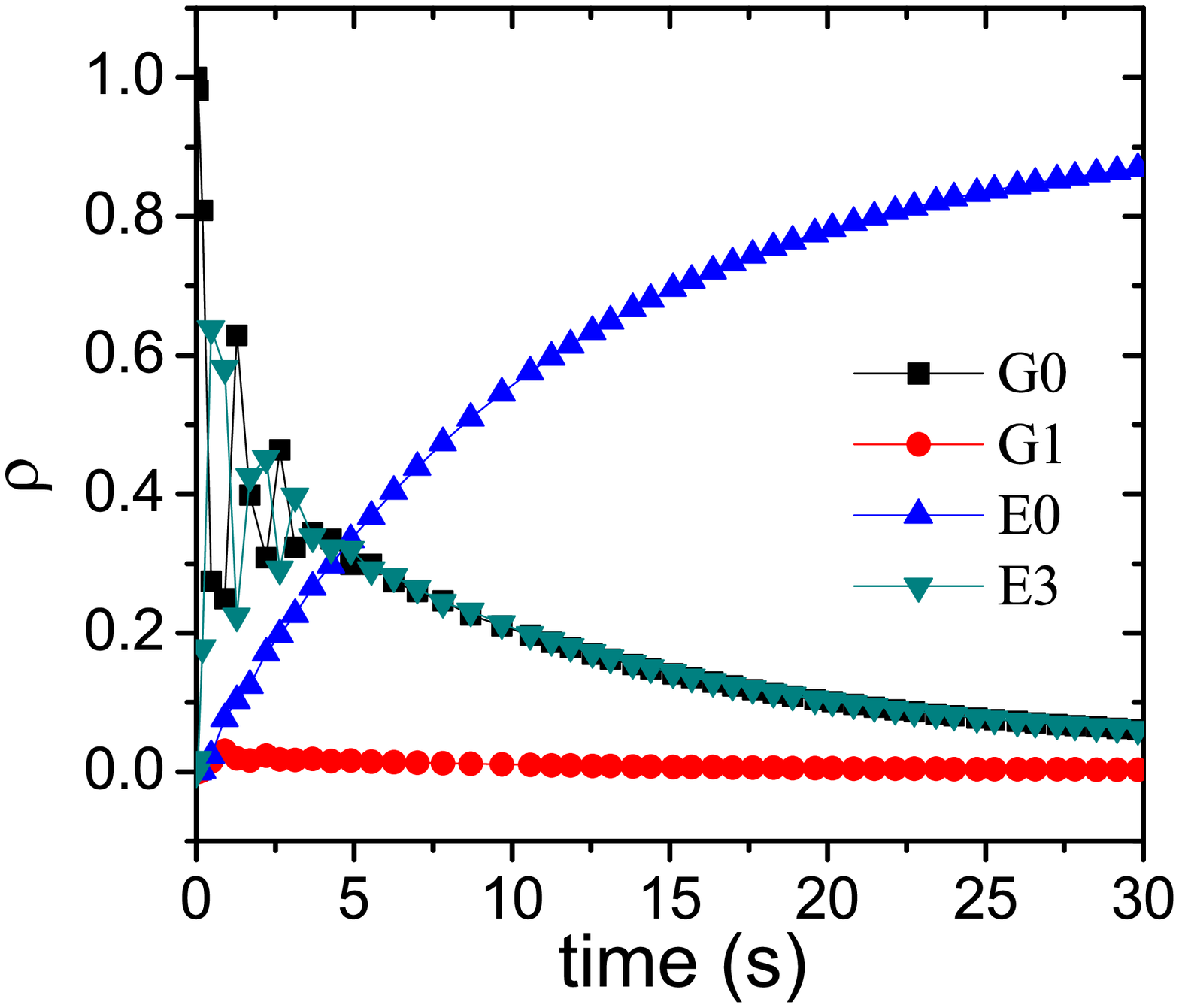}
}
\caption{\label{EXP}
(a) The schematic level diagram for the Nd$^{9+}$ clock. (b) The steady atomic population of the four atomic states for the superradiant lasing of the $\rm{E0}:(5p4f^2)^o_{9/2}-\rm{G1}:(5p^24f)^o_{7/2}$ used for active clock transition of Nd$^{9+}$.}
\end{figure}

Novel applications of optical transitions in highly charged ions (HCIs) for frequency metrology and tests for variation of fundamental constants have made them a subject of much interest lately \cite{Schiller-PRL-2007,Berengut-PRL-2010,Yudin-PRL-2014,Safronova-PRL-2014,Yu-PRA-2016,Yu-PRA-2018,Kozlov-RMP-2018}. The rich energy configuration of HCIs offers numerous optical transitions between the ground state and the long-lived excited states, which can be used for developing high accuracy clocks beyond the 10$^{-18}$ precision limit. The compact sizes of the HCIs make them insensitive to external fields. Strong relativistic effects and high ionization energies make the HCI clock transitions highly sensitive to variation of the fine structure constant $\alpha_e$ and dark matter searches. The recent advances in producing, cooling, and trapping of HCIs using state-of-the-art experimental techniques \cite{Micke-Nature-2020,Liang-PRA-2021} have opened up ways of investigating several ions for developing high accuracy clocks \cite{Windberger-PRL-2015,Nakajima-NIMPS-2017,Bekker-NC-2019}. However, the long-standing concern about the single-ion optical clock is its low stability rate which depends on spectral line widths. The spectral linewidth of the clock transitions in passive-type HCI clocks is limited by the linewidth of the cavity-stabilized laser used as a local oscillator.  The best available lasers have currently been able to attain a mHz-linewidth  \cite{Kessler-NP-2012,Matei-PRL-2017,Jin-APL-2019}, which makes it difficult for the passive-type HCI clocks to beat the low stability limit of the present clocks. The best reported passive-type ion clock, like Al$^+$ ~\cite{Brewer-PRL-2019}, has stability recorded at $1.2\times10^{-15}/\sqrt{\tau}$ ($\tau$ is the averaging time) thus requires an averaging time of more than weeks to attain $10^{-18}$ frequency precision.

In this letter, we explore the possibility to develop a high-accuracy optical clock using the Nd$^{9+}$ ion based on the active-type clock scheme \cite{Chen2005,YuDS-PRL-2007,Chen2009,Meiser-PRL-2009,Kazakov-PRA-2017,Norcia-PRX-2018,Escudero2021}. The Nd$^{9+}$ active clock scheme has two-fold advantage over the previously proposed HCI clocks. Firstly, a $\upmu$Hz-width clock signal is directly generated from superradiation of an ensemble of Nd$^{9+}$ ions, being insensitive to the cavity noise, and therefore the Nd$^{9+}$ active clock eludes the need for the pre-stabilized lasers serving as oscillators, which provides an elegant solution for better frequency stability for the ion clocks. With the $\upmu$Hz-wide clock signal the proposed clock has the potential to improve the stability of two orders of magnitude over the best optical clocks based on the pre-stabilized lasers \cite{Brewer-PRL-2019,Young2020,Beloy2020}. Secondly, the direct generation of the active clock transition avoids the clock integration, that is always tedious for the passive-type HCI clocks, since efficient state preparation and internal state readout must be accomplished by employing the complex quantum logic spectroscopy due to lack of fast-cycling optical in HCIs.

The early knowledge of the energies of the Nd$^{9+}$ ion are insufficient and contradictory, which makes it difficult to extract any useful information for this ion \cite{NIST-data-base,Berengut-PRA-2012,Huo-CJP-2017}. To determine the electronic structure of the Nd$^{9+}$ ion accurately, we have conducted a large-scaled computation by considering the Nd$^{9+}$ ion either as a three-valence electron system with the ground configuration $5p^24f$ above the closed core $[1s^2, . . . , 4d^{10}5s^2]$ or five-valence electron system with the ground configuration $5s^25p^24f$ above the closed core $[1s^2, . . . , 4d^{10}]$. The influence of various factors such as the choice of the basis set and the electronic correlation hierarchy as well as the inclusion of the Breit and QED interactions on the computational results have been taken into account, which ensures us to present the accurate recommended values (see the supplementary materials (SM)). Table~\ref{AtomicStruc} summarizes the energies for the first 2 fine-structure levels of the ground $5p^2~4f$ state and the first 10 fine-structure levels of the excited $5p~4f^2$ state obtained by the multi-reference configuration interaction (MRCI) method \cite{DIRAC}. Our calculations reveal that the energy separation for the two fine-structure levels of the ground state viz $(5p^2~4f)^o_{5/2}$ and $(5p^2~4f)^o_{7/2}$, labeled as G0 and G1, respectively, is 6681 cm$^{-1}$, which is consistent with the earlier calculations \cite{Berengut-PRA-2012,Huo-CJP-2017}. Subsequently, the energy orbital crossing of $4f$ and $5p$ states creates the excited $(5p4f^2)$ configuration that has many fine-structure levels. The first 10 of them are labeled as E0 to E9. Among these levels the excited state $(5p4f^2)^o_{9/2}$ has a rather long lifetime, around 154 seconds. In table~\ref{AtomicStruc} we have also given the hyperfine structure constants $A$, the electric quadrupole moment $\Theta$, the relativistic frequency shift $q$ due to time variation of $\alpha_e$, and the electronic $g$ factor, which are useful for designing a clock. The results in Table~\ref{AtomicStruc} have been verified by adopting another independent atomic structure calculation software AMBiT that is based on the CI + many-body perturbation theory (CI+MBPT) \cite{AMBiT}. Excellent consistencies between the MRCI and AMBiT results are shown in Table III and IV in SM, which validates that our results for the atomic properties of the Nd$^{+}$ are reliable.

\begin{table}[t]
\caption{Energies (EE), lifetime $\tau$, hyperfine structure constant $A$, electric quadrupole moment $\Theta$, and relativistic coefficient $q$ for variation of fine-structure constant, and Land\'{e} $g_j$ factor of the atomic states in the Nd$^{9+}$ ion calculated using MRCI method. \label{AtomicStruc}}
{\setlength{\tabcolsep}{2pt}
\begin{tabular}{lcc ccc c}\hline\hline
Level	&	EE (cm$^{-1}$)	&	$\tau$ (s)	&	$A $ 	&	$\Theta$	&	$q$ 	&	$g_j$	\\
	&	 (cm$^{-1}$)	&	 (s)	&	(MHz)	&	(a.u.)	&	 (cm$^{-1}$)	&		\\ \hline
G0: $(5p^2 4f)^o_{5/2}$	&	0 	&		&	-376 	&	0.019 	&		&	0.830 	\\
G1: $(5p^2 4f)^o_{7/2}$	&	6681 	&	0.23 	&	-175 	&	-0.044 	&	28462 	&	1.156 	\\
E0: $(5p 4f^2)^o_{9/2}$	&	23199 	&	154 	&	-1086 	&	-0.103 	&	441457 	&	0.808 	\\
E1: $(5p 4f^2)^o_{7/2}$	&	24982 	&	0.08 	&	2 	&	0.082 	&	402549 	&	0.829 	\\
E2: $(5p 4f^2)^o_{11/2}$&	29738 	&	0.16 	&	-854 	&	0.076 	&	473150 	&	1.011 	\\
E3: $(5p 4f^2)^o_{9/2}$	&	29680 	&	0.65	&	192 	&	-0.098 	&	470982 	&	1.040 	\\
E4: $(5p 4f^2)^o_{5/2}$	&	30592 	&	0.05 	&	-1352 	&	0.032 	&	433796 	&	0.676 	\\
E5: $(5p 4f^2)^o_{3/2}$	&	30284 	&	0.12 	&	1215 	&	-0.034 	&	411632 	&	0.726 	\\
E6: $(5p 4f^2)^o_{7/2}$	&	32400 	&	0.03 	&	-437 	&	-0.101 	&	468998 	&	1.018 	\\
E7: $(5p 4f^2)^o_{11/2}$&	32309 	&	0.66 	&	307 	&	0.091 	&	464518 	&	1.187 	\\
E8: $(5p 4f^2)^o_{5/2}$	&	33091 	&	0.03 	&	502 	&	0.079 	&	497006 	&	1.152 	\\
E9: $(5p 4f^2)^o_{13/2}$&	34637 	&	0.18 	&	-748 	&	-0.040 	&	418591 	&	1.144 	\\\hline\hline
\end{tabular}}
\end{table}

The working of the active clock is based on a four-level scheme, as illustrated in Fig.~\ref{EXP} (a). A laser around 337 nm is used to pump the Nd$^{9+}$ ions from the ground state G0 to the excited state E3. The ions from the E3 state decay spontaneously down to the E0 state through the M1 transition. Since the E0 state is the most long-lived, more than 85\% of the atomic population is accumulated in this state, leading to a favorable population inversion condition between the E0 state and the G1 state, as illustrated by the steady-state atomic population of the four-level scheme shown in Fig.~\ref{EXP} (b). At this moment a cavity mode has to be adjusted to the E0$\rightarrow$G1 resonance frequency, which makes a so-called `bad-cavity' laser mode \cite{Chen2005,YuDS-PRL-2007,Kuppens-PRL-1994}, so that a self-sustained lasing transition is stimulated between the E0 and G1 states. The ions subsequently decay back to the G0 state through the M1 transition thus accomplish a continuous loop.  An advantage of using this four-level scheme is that the clock states are not obstructed by the light shifts caused by the 337nm pumping laser.

The atomic population of the four-level scheme is calculated by using the Liouville equation of motion without coupling to the cavity \cite{MarIan-quantum optics},
\begin{eqnarray}
\frac{{d \rho_{AP} }}{{dt}} = \frac{1}{{i \hbar }} \left[ {H_{AP}, \rho_{AP} } \right] - \frac{1}{2}\left\{ {\Gamma , \rho_{AP} } \right\}+\Lambda \label{eq1}
\end{eqnarray}
where $\rho_{AP}$ is the density matrix. The first term on the right hand side of Eq. (\ref{eq1}) describes the interaction between the atomic system and the pumping laser with Hamiltonian defined as $H_{AP}=H^0+H^I$, where $H^0= \sum_{n} \hbar\omega_n | J_n \rangle \langle J_n  |$, here $J_n$ represents the six involved levels with $n$ corresponding to G0-G1 and E0-E3, and $H^I= - \hbar\Omega (| J_{G0}\rangle \langle J_{E3}| + |J_{E3}\rangle \langle J_{G0}| )$, $\hbar\omega_n$ is the eigenenergy of $| J_{n}\rangle$,  and $\Omega$ is the pumping Rabi frequency. The second and third terms on right hand side of Eq.\ref{eq1} describe the effects of relaxation and repopulation of the density matrix. The obtained population values for the G0, G1, E0 and E3 states are shown in Fig. 1(b) for pumping Rabi frequency $\Omega=0.5$ Hz. The population values for the E1 and E2 states are nearly zero, thereby not shown in Fig. 1(b).

\begin{table*}[t]
\caption{The sensitivity $K$ to variation of fine structure constant and the sensibility $\langle J \|T^{(2)}\|J\rangle$ (a.u.) to violation of the local Lorentz invariance.  \label{KQnumber}}
{\setlength{\tabcolsep}{16pt}
\begin{tabular}{ccc ccc c}\hline\hline
Ions	&	\multicolumn{2}{c}{$K$}			&&	\multicolumn{2}{c}{$ \langle J \|T^{(2)}\|J\rangle  $}			&	Refs.	\\ \cline{2-3} \cline{5-6}
	&	Transition	&	Value	&&	States	&	Value	&		\\ \hline
Nd$^{9+}$	&	$ (5p4f^2)^o_{9/2} -  (5p^24f)^o_{7/2} $	&	11.37	&&	$(5p^24f)^o_{7/2}$	&	-85.23 	&	This work	\\
	&		&		&&	$(5p4f^2)^o_{9/2} $	&	-85.20 	&		\\[+2ex]
Pr$^{9+}$	&	$5p^2~^3D_2-5p4f~^3G_3$	&	6.32	&&	$5p4f~^3G_3$	&	74.2	&	\cite{Bekker-NC-2019}	\\
	&	$5p^2~^3D_2-5p4f~^3F_2$	&	5.28	&&	$5p4f~^3F_2$	&	57.8	&		\\[+2ex]
Yb$^{+}$	&	$4f^{14}6s~^2S_{1/2}-4f^{13}6s~^2F_{7/2}$	&	-5.95	&&	$4f^{14}6s~^2D_{3/2}$	&	9.96	&	\cite{Flambaum-CJP-2009,Dzuba-NP-2016}	\\
	&		&		&&	$4f^{14}6s~^2D_{5/2}$	&	12.08	&		\\
	&		&		&&	$4f^{13}6s~^2F_{7/2}$	&	-135.2	&		\\[+2ex]
Ca$^{+}$	&	$4s~^2S_{1/2}-3d~^2D_{3/2}$	&	0.07	&&	$3d~^2D_{3/2}$	&	7.09 (12)	&	\cite{Flambaum-CJP-2009,Pruttivarasin-Nature-2015}	\\
	&	$4s~^2S_{1/2}-3d~^2D_{5/2}$	&	0.08	&&	$3d~^2D_{5/2}$	&	9.25 (15)	&		\\ \hline\hline
\end{tabular}}
\end{table*}

\begin{figure}[t]
\centering
 \subfigure[ ]{
\includegraphics[width=7cm,height=5.5cm]{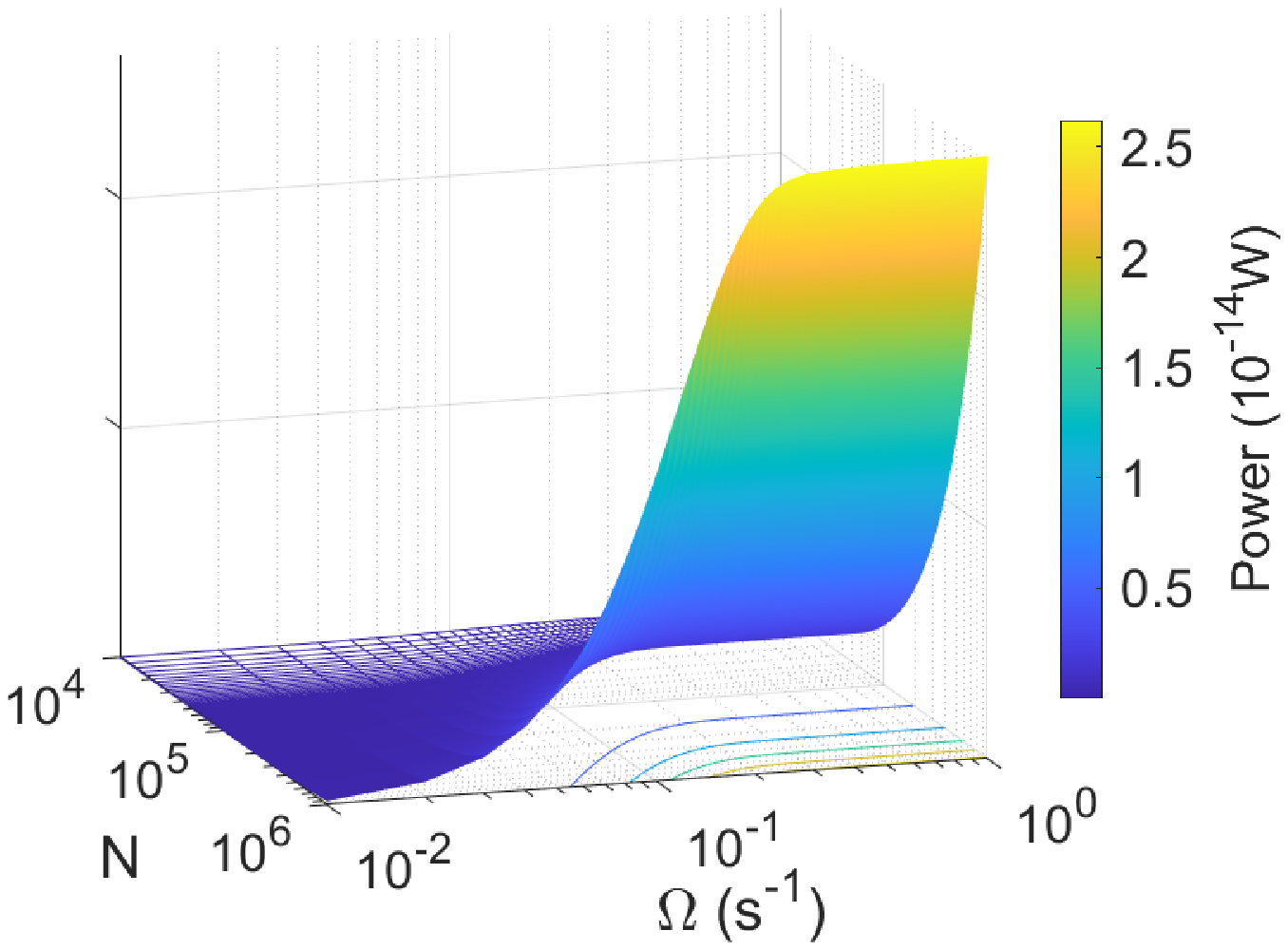}
}
\subfigure[ ]{
\includegraphics[width=7cm,height=5.5cm]{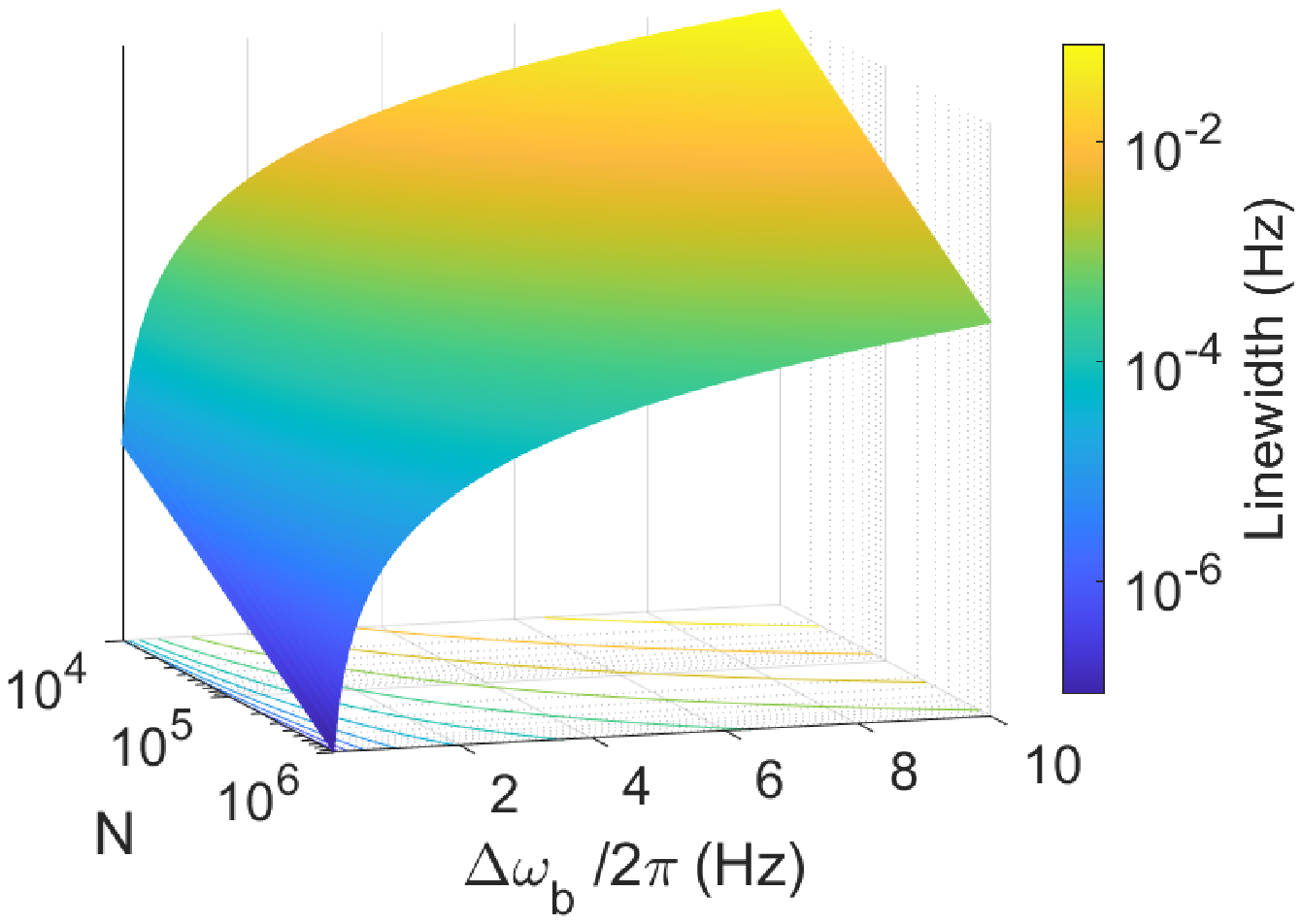}
}
\caption{\label{fig2}
(a) Power as a function of pump Rabi frequency $\Omega$ and ion number N. (b) Linewidth of the optical clock as a function of inhomogeneous broadening $\Delta \omega_b$ and ion number N. The cavity dissipation rate is set to be $\kappa=2\pi\times10$kHz for both subfigures.}
\end{figure}

Correspondingly, when operated in the bad cavity mode the output power of the superradiant lasing in the atom-cavity coupling regime is evaluated by solving the Liouville equation for the density matrix $\rho_{AC}$,
\begin{eqnarray}
\frac{{d \rho_{AC}} }{{dt}} = \frac{1}{{i \hbar }} \left[ {H_{AC}, \rho_{AC} } \right] - \frac{1}{2}\left\{ {\Gamma , \rho_{AC} } \right\}+\Lambda.\label{eq4}
\end{eqnarray}
The total Hamiltonian in this case is given by $H_{AC}=H^0+H^I+H^c$, where the Hamiltonian for interaction between ions and intra-cavity stimulated laser field $H^c=-\hbar g\sqrt{n_p+1}( |J_{G1}\rangle \langle J_{E0}| + |J_{E0}\rangle \langle J_{G1}| )$, with the atom-cavity coupling constant $g= \mu \sqrt {\frac{2\pi\omega_a }{{\hbar {\varepsilon _0}V_c}}} = $ 16 Hz for the transition matrix element $\mu$ between E0 and G1 states, the cavity mode volume $V_c=1\times10^{-11}\rm {m^3}$, $\omega_a$ is the angular frequency of the superadiation lasing, and $\varepsilon_0$ is the permittivity of vacuum.  Dimensionless intensity $n_p$ defines the number of photons inside the cavity, which satisfies the following equation
\begin{eqnarray}
\frac{{dn_p}}{{dt}} = ig\sqrt{n_p+1} (\rho^{G1E0}_{AC}-\rho^{E0G1}_{AC})-\kappa n_p,\label{eq5}
\end{eqnarray}
with $\rho^{G1E0}_{AC}$ and $\rho^{E0G1}_{AC}$ corresponding to the off-diagonal element of the density matrix $\rho_{AC}$, and $\kappa=2\pi\times10$ kHz being the cavity dissipation rate. The output power of the superradiant lasing is estimated as $P = N\hbar {\omega _a}\kappa n_{s}$, where $n_{s}$ is the steady-state solution of $n_{p}$, calculated by Eqs. (\ref{eq4}) and (\ref{eq5}). The output power as a function of the pumping Rabi frequency $\Omega$ and the ion number N is shown in Fig. \ref{fig2} (a). Given the number of the trapped ions $N=10^6$ and $\Omega=0.5$ Hz we estimate the total steady-state photon number inside the cavity to be $n_t=n_s N=1.3$, and the output power of the superradiant lasing to be $P=2.3 \times 10^{-14}$ $\rm{W}$.

The linewidth of the optical clock based on inhomogeneous broadening of an ensemble of ions can be formulated as \cite{Khoury-RPA-1996},
\begin{equation}
\label{eq:linewidth}
\Delta \nu =[1+\frac{(1-b)\Gamma_{G1}+2(1+b)\Gamma_{E0}}{4(\Gamma_{G1}+\Gamma_{E0})}I]
(\frac{\Gamma_{E0G1}}{\Gamma_{E0G1}+b\kappa})^2\frac{\kappa}{2n_t},
\end{equation}
where $\Gamma_{G1}$ and $\Gamma_{E0}$ are the total decay rates from the two clock states, G1 and E0, respectively, $\Gamma_{E0G1}$ is the decay rate from E0 to G1, and $b$ and $I$ are the coefficients that describe the inhomogeneous broadening and the power broadening, respectively, with
\begin{equation}
\label{eq:b}
b=\frac{2}{\sqrt{\pi}}\frac{\beta}{\alpha}\frac{ \rm{exp}[-\beta^2/\alpha^2]}{\rm{erfc}(\beta/\alpha)}-\frac{2\beta^2}{\alpha^2},
\end{equation}
where $\beta=\sqrt{1+I}$, $\alpha=\triangle \omega_b/\Gamma_{E0G1}$, and $\Delta \omega_b$ is the inhomogeneous broadening. When $\Delta \omega_b$=0 Eq. \ref{eq:b} transforms to the quantum limited linewidth of a bad cavity laser \cite{KuppenLinewidth}. The power broadening is determined by $I=n_t/n_{sat}$, wherein the saturation photon number $n_{sat}$ is determined by $n_{sat}=\Gamma_{E0G1}\Gamma_{G1}\Gamma_{E0}/(2g^2(\Gamma_{G1}+\Gamma_{E0}))$. For cavity dissipation rate $\kappa=2\pi\times10$ kHz, we obtain linewidth of the optical clock as a function of $\Delta \omega_b$ and $N$ shown in Fig. \ref{fig2}(b). Given the total steady-state photon number inside the cavity $n_t=1.3$, and $\Delta \omega_b$ equal to empirical values ranging from $2\pi \times 0.5$ Hz to $2\pi \times 10$ Hz\cite{Champenois-PRA-2010}, the linewidth $\Delta \nu$ of the supperradiant lasing can be estimated within 2 $\upmu$Hz to 800 $\upmu$Hz.

Table \ref{KQnumber} summarizes the sensitivity of the E0: $(5p4f^2)^o_{9/2}$$-$G1: $(5p^24f)^o_{7/2} $ transition to variation of $\alpha_e$ and violation of the lorentz invariance. The relativistic energy shift due to the variation of $\alpha$ is defined in terms of angular frequency of  the transition $\omega_t$ corresponding to the value of fine structure constant $\alpha^t_e$ at time $t$ such that $x=(\alpha^t_e/\alpha^0_e)^2-1$ and
\begin{equation}
\omega_t=\omega_0+qx.
\end{equation}
Here $\omega_0$ is the angular frequency of the transition for the present-day value of the fine-structure constant $\alpha^0_e$. The relative sensitivity coefficient $K$ is defined by $K=2q/\omega_0$. The $K$ value for the $ \rm{E0}:(5p4f^2)^o_{9/2}-\rm{G1}:(5p^24f)^o_{7/2} $ transition in Nd$^{9+}$ surpass the $K$ values for transitions in some of other clock candidates such as the $5p^2~^3D_2-5p4f~^3F_2$ and $5p^2~^3D_2-5p4f~^3F_2$ transitions in Pr$^{9+}$ and the $4f^{14}6s~^2S_{1/2}-4f^{13}6s^2~^2F_{7/2}$ transition in Yb$^{+}$. The sensitivity to invariance under local Lorentz transformation is formulated using  operator $T^{(2)}_0$, defined as
\begin{equation}
T^{(2)}_0=c\gamma_0(  {\bf \gamma} \cdot {\bf p}-3\gamma_z p_z),
\end{equation}
where $c$ is the speed of light, $(\gamma_0, {\bf \gamma})$ are Dirac matrices, and ${\bf p}$ is the momentum of a bound electron. We find that the matrix elements $\langle J \|T^{(2)}\|J\rangle$ for the $\rm{G1}:(5p^24f)^o_{7/2}$, and $\rm{E0}:(5p4f^2)^o_{9/2}$ states are comparable to the matrix elements for the clock transition states for Pr$^{+}$ and Yb$^{+}$ clocks. We have also examined the effect of  external fields and their gradients on the clock transition, as given in Table VIII in SM, which supports an accuracy of the proposed clock to $10^{-19}$ level.

The Nd element has rich natural isotopes, and its ionization energy being $\sim$152(8)eV is relatively low, \cite{NIST-data-base}, therefore the Nd$^{9+}$ ion should be easily produced in our home-build SW-EBIT (Shanghai-Wuhan Electron Beam Ion Trap) whose electron beam energy typically ranges from tens of eV to keV\cite{Liang-RSI}. High sympathetic cooling efficiency of Nd$^{9+}$ is possible by utilizing Mg$^{+}$ ions with a mass ($M$) to charge ($Q$) ratio of $(M/Q)_{Nd}:(M/Q)_{Mg} = 0.7:1$. In Fig. ~\ref{EXP} (a) we plot also an 431nm transition between the ground state and the E0 state that is suitable for making the passive-type clock. For the passive Nd$^{9+}$ clock, the quantum logic-assisted electronic state preparation and the clock interrogation can utilize the unstable excited states in Nd$^{9+}$, such as E4: $(5p 4f^2)^o_{5/2}$, E6: $(5p 4f^2)^o_{7/2}$, E8: $(5p 4f^2)^o_{5/2}$, and another ground-state fine structure splitting G2: $(5p^2 4f)^o_{5/2}$ (see Table III in SM). These states have relatively fast decay rate and nearly 100\% branch ratio to the ground state (see Table V and VI in SM), and are compatible with the Mg$^{+}$ ion assisting the quantum logic operation based on the Nd$^{9+}-$Mg$^+$ ion pair.

In conclusion, we propose the Nd$^{9+}$ ion as a potential candidate of $\upmu$Hz linewidth optical clocks based on the E0: $(5p4f^2)^o_{9/2}$$-$G1: $(5p^24f)^o_{7/2} $ active clock transition. The Nd$^{9+}$ ion shares common advantages of other HCI clock candidates such as low systematic uncertainty to external perturbations and high values of $K$ and $\langle J \|T^{(2)}\|J\rangle$. More importantly, the active clock based on the collectively self-sustained emission of an ensemble of the Nd$^{9+}$ ions in a bad cavity can solve the constraint of low stability limitation of the passive-type HCI clocks. Like Hydrogen maser, the Nd$^{9+}$ active clock is favorable for its excellent short-term stability. Furthermore, with linewidth decreasing two orders of magnitude less than the current best mHz-linewidth laser, the Nd$^{9+}$ active clock presents potentially prominent long-term instability being $10^{-18}\sim10^{-20}$ for $10^3\sim10^{6}$ seconds averaging time (pattern is shown in Fig. 1 of SM). The frequency pulling \cite{Audoin1981} due to the cavity drift during a such long averaging time could be suppressed by synchronizing the frequency of the cavity mode with another atomic frequency reference.

We thank B. K. Sahoo for help and discussions. The work was supported by the Strategic Priority Research Program of the Chinese Academy of Sciences (CAS), Grant No. XDB21030300, National Natural Science Foundation of China(NSFC)(No.91436210, No.11874064, No.11934014, No.12004392) and the National Key Research and Development Program of China, and the NKRD Program of China, Grant No. 2016YFA0302104. The work of B.A. is supported by SERB-TARE(TAR/2020/000189), New Delhi, India.

\end{document}